\documentstyle[preprint,aps,tighten]{revtex}

\begin{document}
\draft
\title{Two programs for studying stellar evolution and nuclear astrophysics}
\author{T. E. Liolios$^{1}$ \thanks{%
www.liolios.info}}
\address{Hellenic Merchant Navy Academy, Department of Science\\
Hydra Island 18040, Greece}
\maketitle

\begin{abstract}
Two FORTRAN programs are presented which plot the Hertzsprung-Russell
diagram and the temporal evolution of such stellar quantities as: central
and photospheric isotopic abundances, central densities and temperatures,
luminosities,effective temperatures and photospheric radii for a wide range
of stellar masses. The programs, which are modifications and extensions of
some modules of the {\it TYCHO 6.0} stellar evolution package, are
accompanied by various control input files as well as by a library of data.
The data library is actually one of the output files generated by {\it TYCHO
7.0} (a modified version of {\it TYCHO 6.0}), while the plots themselves are
generated by the PGPLOT Graphics Subroutine Library which is also publicly
available and well cited.
\end{abstract}

\section{Introduction}

{\it TYCHO 6.0} \cite{tycho60} is a general, one dimensional (spherically
symmetric) stellar evolution code, designed for hydrostatic and hydrodynamic
stages, using state of the art procedures and microphysics. It is written in
structured FORTRAN (f77) and has extensive on-line graphics using the PGPLOT
graphics package\cite{pgplot} in X-windows environments. It has being
developed as an open-source,community code and can run very effectively on
Linux laptop and desktop computers. {\it TYCHO 6.0} is being used by the
University of Arizona astrophysics group (as well as by the author) for
ongoing research in stellar evolution (for relevant papers see \cite{tycho60}
) while it has also been used to generate low-noise initial models for
multi-dimensional numerical hydrodynamic simulations. The {\it TYCHO 6.0}
package includes various auxiliary files and subprograms which can monitor
both the evolution of a star and nuclear reaction networks under various
conditions.

Although {\it TYCHO 6.0's} auxiliary programs focus on various aspects of
stellar evolution, such as the Hertzsprung-Russell (HR) diagram or the
composition and thermodynamics of each evolutionary sequence, there is no
auxiliary program which can investigate the temporal evolution of the star
(neither during nor after the actual simulation process). Since typical HR
diagrams and static composition profiles are actually bereft of any temporal
information reliable programs are needed to satisfy this obvious need for
all {\it TYCHO 6.0} users. Moreover, such programs should not just be able
to be used by TYCHO users. Anyone wishing to study the evolution of a star
could benefit from an independent program which could be applied to a large
library of data generated by TYCHO. This is the purpose of the programs
presented here.

In the sections that follow we first describe the programs and then its
input and output files. Then we implement the programs by performing a
test-run for a 3M star and a particular initial composition. The present
paper initiates an effort to improve the {\it TYCHO 6.0} package and extend
its capabilities in order to investigate various nuclear astrophysics
topics. A similar effort is going on by the university of Arizona
astrophysics group which has already derived a new version, called {\it %
TYCHO 6.11}, which is also publicly available\cite{tycho60}. {\it TYCHO 6.0}
has now been equipped by the author with screening effects in the derivation
of thermonuclear reaction rates as well as with various new auxiliary
programs such as those presented in this paper. Screening corrections in the
equation of state for completely degenerate environments (e.g. White-Dwarf
progenitors) will soon be available These aspects are not included in the 
{\it TYCHO 6.11} version which, however, has very interesting improvements
over {\it TYCHO 6.0}. To facilitate versioning and differentiation from the 
{\it TYCHO 6.11} sequence (and all relevant work from the Univ.of Arizona
group), the version independently evolved and implemented here will be
referred to as {\it TYCHO 7.0}, while all relevant improvements and
modifications over the {\it TYCHO 6.0} version will be consistently recorded
and published, starting with the present paper. Note that whenever the
stellar evolution code is referred to as {\it TYCHO} then the relevant
comment is applicable to all versions of the code.

\section{HRTEMP}

\subsection{Brief description}

{\it Hrtemp} is an {\it F77} program which reads a library of data ({\it %
hr.prefix}, generated by {\it TYCHO 7.0}) as well as some control input
files in order to plot the temporal evolution of such stellar quantities as:
central and photospheric isotopic abundances, central densities and
temperatures, luminosities,effective temperatures and photospheric radii for
a wide range of stellar masses. The program makes use of the PGPLOT Graphics
Subroutine Library which is also publicly available and well documented. The
source file {\it hrtemp.f} is located in {\it /hr/src} along with its {\it %
Makefile.Linux} and {\it Makefile} files, which are actually those
distributed with the {\it TYCHO 6.0} package with minor modifications. Note
that the {\it Makefile.Linux} file should be informed about the location of
the files {\it libpglot.a, libX11.a} on the user's machine, while the PGPLOT
package should have been already installed.

\subsection{Input files}

\subsubsection{hr.prefix}

While simulating the evolution of a star, TYCHO generates at each time step
a grid of data which include such quantities as : time, central and
photospheric isotopic abundances, central densities and temperatures,
luminosities,effective temperatures, mass loss/gain, angular rotational
velocity and photospheric radii. All these data are stored in a file which
is used by the program {\it hrtemp} presented here. This file is denoted by 
{\it hr.prefix}, where {\it prefix} is a two-character identification of the
stellar evolution event. For example {\it hr.n1} indicates the evolutionary
sequence of a 1M$_{\odot }$ star, {\it hr.n2} that of a 2M$_{\odot }$ star
and so on. Three such files are included in the present work. All
evolutionary points in {\it hr.prefix} are numbered thus one can easily
realize the density of the mesh. The size of such files is roughly 5MB to
10MB.

The {\it hrtemp} program can be integrated into the {\it TYCHO 7.0} package
by all its users in order to extend its capabilities. Of course it can
always be used independently by simply using a library of pregenerated {\it %
hr.prefix} files. A typical form of an {\it hr.prefix} file is an array of
the following blocks, each of which corresponding to a particular time-step:

{\it model serial number, iterations, time,timestep, photospheric radius,
photospheric density, photospheric velocity, log(L/Lo), log(Teff), mass
interior, mass loss per year, solar masses ejected, angular rotational
velocity, central temperature, central density,photospheric abundances
(array), central abundances(array).}

\subsubsection{The control file hrtemp.in}

A control input file {\it hrtemp.in} is essential to the implementation of
the program. Its form is as follows:\newline

{\it '.....................................................................'%
\newline
' HRTEMP: for TYCHO-7.0 '\newline
'.....................................................................'%
\newline
' Display device: /ps, /xwin ................' 'device' '/xwin'\newline
' Number of sequences to plot................' 'nfiles' 3\newline
' Filename of sequence.......................' 'hrfile' 'hr.n1'\newline
' Filename of sequence.......................' 'hrfile' 'hr.n3'\newline
' Filename of sequence.......................' 'hrfile' 'hr.n5'\newline
'log(L/L}$_{\odot }${\it )=0,logTe=1,log(R/R}$_{\odot }${\it %
)=2,Xej=3,Xc=4,log }${\it T}_{c}${\it =5,log }$\rho _{c}${\it =6: ifunct'%
\newline
' lg(L/Lo),lgTe,lg(R/R}o{\it ),Xej,Xc,lgTc,lgDc= ' 'ifunct' 4\newline
' The serial number of the isotope (net.rc)..' 'isotop' 79\newline
' minimum log(t) for x axis..................' 'xmin' 10.9\newline
' maximum log(t) for x axis..................' 'xmax' 18.79\newline
' minimum Yi for y axis.....................' 'ymin' 0.0d0\newline
' maximum Yi for y axis.....................' 'ymax' 1.0d0\newline
' Line style.1=solid line, else dots.........' 'linesty' 1\newline
'.....................................................................'%
\newline
}

The display device is defined in the fourth line. There are three
options:/ps,/cps,and /xwin. The first two generate a figure in the form of a
postscript file (named pgplot.ps) either black and white (/ps) or
color(/cps), while the /xwin option generates an on-screen figure.

The fifth line indicates the number of sequences to be plotted in the same
figure. Different plots are drawn using different line styles (not available
in {\it TYCHO 6.0}) and colors. Thus, for example, one can use the input
files {\it hr.n1,hr.n3} to follow the simultaneous temporal evolution of two
stars (1M$_{\odot }$ and 3M$_{\odot }$).

The sixth line (and all the similar ones that have to follow if the number
of sequences specified in the fifth line is larger than unity) indicate the
data files to be used. If only one sequence is to be followed then only one
such line is necessary.

The seventh line is a dummy text line which serves as a reminder of the
meaning of the switch to be used in the eighth line.

The eighth line stands for the switch {\it ifunct}, which defines the
quantity to be plotted versus time. Namely: 
\[
log(L/L_{\odot })=0,logTe=1,log(R/R_{\odot })=2,X_{ej}=3,X_{c}=4,log\left(
T_{c}\right) =5,log\left( \rho _{c}\right) =6 
\]
where $X_{ej}(X_{c})$ is the isotopic abundances of the ejecta (in the
center), $T_{c},\rho _{c}$ are the central temperatures and densities
respectively,$T_{e}$ is the effective temperature, $L$ is the total
luminosity, and $R$ is the photospheric radii. The serial number of the
isotope to be plotted (if the {\it ifunct} option is 4 or 3) is defined in
the ninth line. That number can be found in the file {\it net.rc} which is
included in this work (and is identical to the one found in {\it TYCHO 6.0}.
For example 80 stands for $^{4}He$, 79 for $H,$ and so on. That line is
ignored unless the {\it ifunct} quantity is 4 or 3. Note that TYCHO 6.0 uses
a small reaction network for small temperatures ($T<10^{7}K$), which
consists of only 29 isotopes (i.e. the first 26 of {\it net.rc} plus $^{4}He$%
, $H$ and electrons), which should be taken into account when plotting the
temporal evolution of an isotope.

The temporal limits for the plot are defined in the tenth and eleventh
lines. Actually those limits stand for the logarithm of time $log(t)$ when
time is measured in seconds.

The limits for the vertical axes are defined in the twelfth and thirteenth
lines. Note that except for the isotopic abundances all other limits are
measured on a logarithmic scale.

\subsubsection{Other auxiliary files}

The source code makes use of three files to be included upon compilation of
the program namely: {\it dimenfile,cconst,caeps}. {\it Dimenfile} defines
very fundamental quantities such as the dimension of the abundance arrays; 
{\it ccosnt} defines some of the constants to be used;and finally {\it aeps}
defines some of the arrays used in the simulation. Note that these three
files are identical to the files distributed with the TYCHO package and in
order to avoid any confusion we have retain the definition of various other
quantities which are necessary to all {\it TYCHO} packages but are not
needed in the present {\it hrtemp} program.

Another auxiliary file which has already been mentioned is {\it net.rc}
which contains information about the isotopic network which has been used in
the simulation. The data stored in that file are easy to follow and are of
the form:

{\it serial number, protons, neutrons, mass, abundance}

\subsection{Output files}

The main function of the {\it hrtemp} program is plotting the temporal
evolution of various stellar quantities for one or more simulations. {\it %
hrtemp} can generate an on-screen output (which is a family of plots against
a black background) or a postscript file , which is a family of plots
against a white background. The postscript file is named {\it pgplot.ps}.
The quality of the output file, depends of course on the quality of the
simulation itself. Therefore, one should have detailed information about the
assumptions and the approximations entailed during the simulations. The
assumptions made for the derivation of all the {\it hr.prefix} files are the
same as the ones made in $\cite{young}$, except for the mixing length which
in our case is set equal to {\it 1.8}. Although we have equipped {\it TYCHO
7.0} with Mitler's\cite{mitler,mitlerjpg,salpepj} screening effects in the
derivation of thermonuclear reaction rates, for simplicity we have chosen to
turn screening off. A major parameter in the quality of our {\it hr.prefix}
data files is the range of validity of the equation of state. We have run 
{\it TYCHO 7.0} for various initial stellar masses until it cannot converge
any more. However, since TYCHO's EOS needs adjustment for strong screening
effects the quality of the plots in strongly screened domains (e.g.
white-dwarf progenitor phases) may not be very satisfactory. Of course, all
weakly screened stellar interiors are very reliably simulated and the
corresponding plots can be safely used.

Another aspect of this program is that each {\it hr.prefix} file is
generated not only for a specific stellar mass but also for a specific
stellar composition: {\it X,Y,Z}. In fact all {\it hr.prefix} files which
accompany the present program have been generated for a $X=0.7,Y=0.28,Z=0.02$
composition. The users of the {\it hrtemp} program (or the {\it hrplot} one
whose description follows) can always request new {\it hr.prefix} files from
the author, specifying each time the initial composition and stellar mass as
well as all the other details of the requested simulations.

Regarding the quality of the plots, it should be emphasized that there is a
wide range of options such as line width, colors, magnification etc. All
these options can be modified in the source code if the user has some
experience in using PGPLOT. Since the manual of PGPLOT is publicly available
along with the source code itself no further analysis is deemed necessary
here. All the advantages and disadvantages of using PGPLOT are built into
the {\it hrtemp} plots. A thorough discussion of those can be found in the
PGPLOT manual and in all relevant internet sites.

\section{HRPLOT}

\subsection{Description}

While {\it hrtemp} uses the file {\it hr.prefix} to plot the temporal
evolution of the star, there is another auxiliary file distributed freely
with TYCHO 6.0, namely {\it hrplot.f}, which uses the same input file to
plot the HR diagram of the evolution. The {\it TYCHO 7.0} version stores
more data in the {\it hr.prefix} file than {\it TYCHO 6.0} does, such as
central isotopic abundances, central temperatures and densities etc.
Therefore, the original program {\it hrplot.f} has been modified accordingly
to be able to read the {\it hr.prefix} file generated by {\it TYCHO 7.0}.
Moreover, the new version of {\it hrplot} included in the {\it TYCHO 7.0}
package has been further improved so that evolutionary lifetimes can now be
depicted on the HR diagram. Except for those two improvements, all other
aspects of the {\it hrplot} are similar to those of the {\it TYCHO 6.0}
package.

\subsection{Input files}

\subsubsection{hr.prefix}

The major input files used by {\it hrplot} are of course the {\it hr.prefix}
files which have already been adequately analyzed in the description of {\it %
hrtemp}.

\subsubsection{The control file hrplot.in}

A control input file {\it hrplot.in} is essential to the implementation of
the {\it hrplot} program. Its form is as follows:\newline
{\it '.....................................................................'%
\newline
' HRPLOT: for TYCHO-7.0 '\newline
'.....................................................................'%
\newline
' Display device: /ps, /xwin ................' 'device' '/xwin'\newline
' Number of sequences to plot................' 'nfiles' 3\newline
' Filename of sequence.......................' 'hrfile' 'hr.m1'\newline
' Filename of sequence.......................' 'hrfile' 'hr.m3'\newline
' Filename of sequence.......................' 'hrfile' 'hr.m5'\newline
' 0 gives -logTe,logL,else logR,logL.........' 'iradii' 0\newline
' minimum log Xi for x axis..................' 'xmin' 4.2\newline
' maximum log Xi for x axis..................' 'xmax' 3.3\newline
' minimum log Yi for y axis..................' 'ymin' -1.0\newline
' maximum log Yi for y axis..................' 'ymax' 3.5\newline
' Line style.1=solid line, else dots.........' 'linesty' 1\newline
' Some observational points? (0=no)..........' 'iobs' 0\newline
' Evolutionary lifetimes?(0=no else number= ' 'nums' 0\newline
' Evolutionary lifetime (nums}${\it \neq }${\it 0)........' 'logL' 2.028d0%
\newline
' Evolutionary lifetime (nums}${\it \neq }${\it 0)........' 'logT' 4.06d0%
\newline
' Evolutionary lifetime (nums}${\it \neq }${\it 0)........' 'logL' 2.153d0%
\newline
' Evolutionary lifetime (nums}${\it \neq }${\it 0)........' 'logT' 4.029d0%
\newline
' Evolutionary lifetime (nums}${\it \neq }${\it 0)........' 'logL' 1.918d0%
\newline
' Evolutionary lifetime (nums}${\it \neq }${\it 0)........' 'logT' 3.688d0%
\newline
'.....................................................................'%
\newline
}

The first eight lines bear the same meaning as those of {\it hrtemp.in}.

The fifth line carries a switch ({\it iradii}), which defines the type of
plot to be generated by {\it hrplot} and is self-explanatory.

The next four lines are the logarithms of the luminosity and the effective
temperature which will be the limits of the HR diagram.

The line style is defined by the switch linesty in the next line, just as it
is the case with {\it hrtemp.in}.

The program {\it hrplot} included in the {\it TYCHO 6.0} package has the
capability of depicting observational points on the HR diagram and that is
controlled by the switch {\it iobs}. The observational points are defined in
the source code {\it hrplot.f}.

The novel aspect in the {\it hrplot} presented here is the inclusion of
evolutionary lifetimes, which is controlled by the switch {\it nums}. The
user can enable that option defining simultaneously the number of
evolutionary points to be included. The evolutionary points will be shown on
the HR diagram as integers, each of which corresponding to a particular
point $(log(T_{eff}),log(L/L_{\odot }))$ according to the data in {\it %
hrplot.in}.

\subsection{Output files}

The output file is of course an HR diagram whose accuracy is a function of
the accuracy of the simulation itself. It should be emphasized that all {\it %
TYCHO} codes can start the evolution from a homogeneous gas sphere and
follow the evolution until very advanced stages are reached (white-dwarf
progenitor, pre-supernovae etc.). In some cases (low-mass stars) the first
few steps of the simulation may fail to converge but {\it TYCHO} soon
overcomes the problem and converges rapidly along the Hayashi track. This
effect is obvious at the HR plot of a 1M star where there is a small
irregularity very high up the Hayashi track. The user can either discard (or
simply disregard) the first non convergent steps of the {\it hr.prefix} file
or can always choose to truncate a larger portion of the simulation and
start the HR plot from the Zero-Age Main Sequence ({\it ZAMS}). Regarding
the quality of the plots once again we should underline that it depends on
the actual performance of the PGPLOT package.

Note that if the user switches {\it nums} on then the present sun is always
indicated by a star on the HR diagram.

\section{Test run}

We follow the evolution of 1M$_{\odot }$, 3M$_{\odot }$ and 5M$_{\odot }$
stars ($X=0.7,Y=0.28,Z=0.02$) by running {\it TYCHO 7.0} until it fails to
converge. The {\it hr.prefix} files that we have generated are respectively 
{\it hr.n1,hr.n3,hr.n5}. Then we implement our programs by plotting the
temporal evolution of: a) the helium, hydrogen, and carbon-12 central
abundances, b) the luminosity, c) the effective temperatures, d) the central
temperature, e) the central density, and f) the radius, for a 3M$_{\odot }$
star. We also plot the HR diagram for the same star including some
enumerated life-times.

After $log(t)=16$ the star begins its ascend up the Asymptotic Giant Branch
undergoing various breathing pulses. Those pulses are indicated by the
irregularities which the plots depict in the relevant temporal regime. The
physics of the AGB is being further elaborated in version {\it TYCHO 7.0} by
including strong screening corrections to the EOS (to be published). Of
course, the programs {\it hrtemp} and {\it hrplot} presented here can be
safely used to study stellar evolution before the AGB regime.

{\bf FIGURE CAPTIONS}

Figure 1. The evolutionary track of a 3M$_{\odot }$ star in the H-R diagram.
The time required to reach the enumerated points is given in $hrplot.in$.

Figure 2. The temporal evolution of the central temperature during the
evolution of a 3M$_{\odot }$ star. After $log(t)=16$ the star begins its
ascend up the Asymptotic Giant Branch undergoing various breathing pulses.

Figure 3. The temporal evolution of the central density during the evolution
of a 3M$_{\odot }$ star.

Figure 4. The temporal evolution of the central hydrogen abundance (mass
fraction) during the evolution of a 3M$_{\odot }$ star.

Figure 5. The temporal evolution of the central helium abundance (mass
fraction) during the evolution of a 3M$_{\odot }$ star.

Figure 6. The temporal evolution of the central carbon-12 abundance (mass
fraction) during the evolution of a 3M$_{\odot }$ star. The comment in
Figure 5 abou AtGB pulses applies here as well.

Figure 7. The temporal evolution of the total luminosity during the
evolution of a 3M$_{\odot }$ star.

Figure 8. The temporal evolution of the stellar radius during the evolution
of a 3M$_{\odot }$ star.

{\bf ACKNOWLEDGMENTS}\newline
The author is grateful to Prof. D.Arnett for making TYCHO available to all
of us as an open source stellar evolution code.

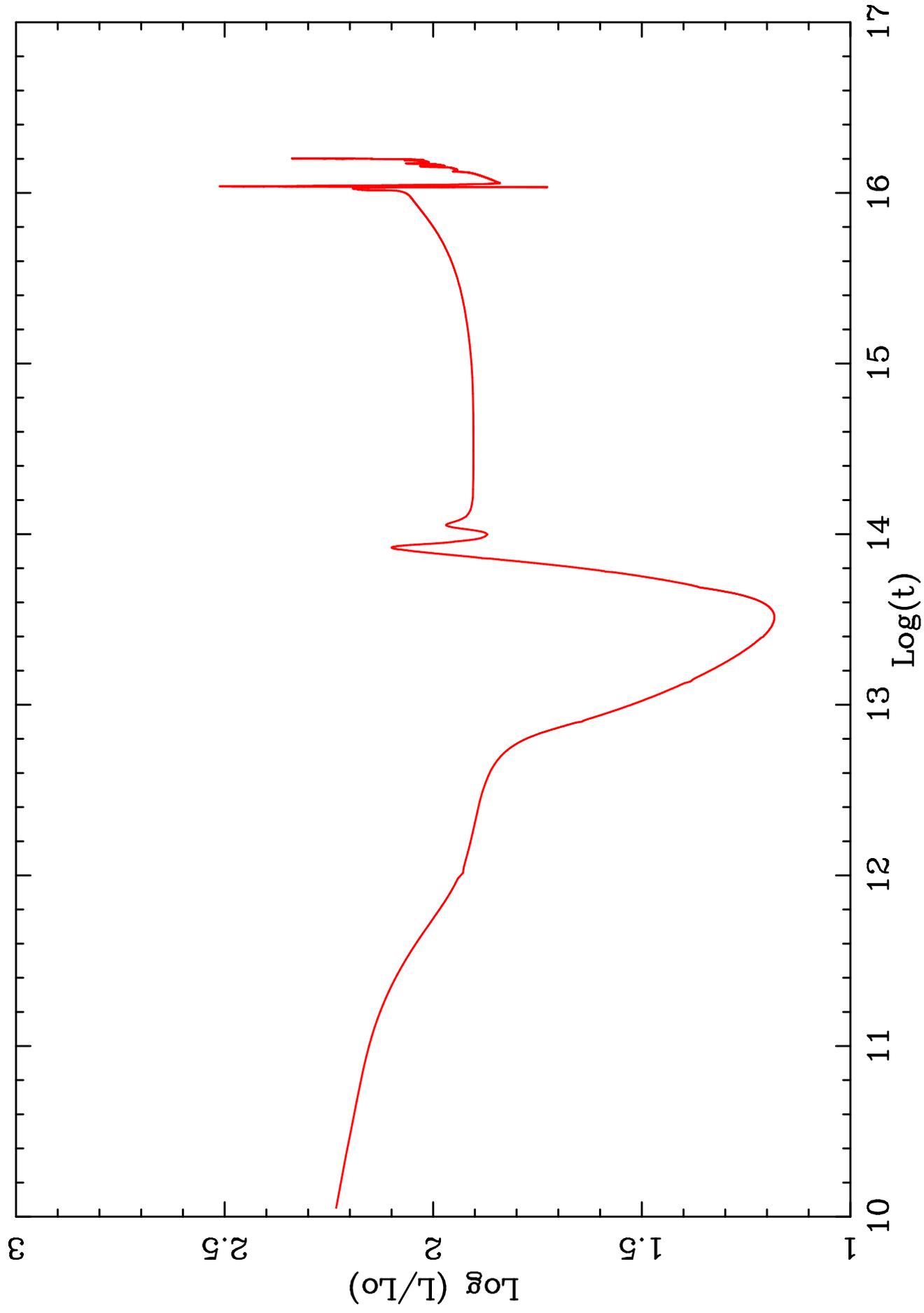
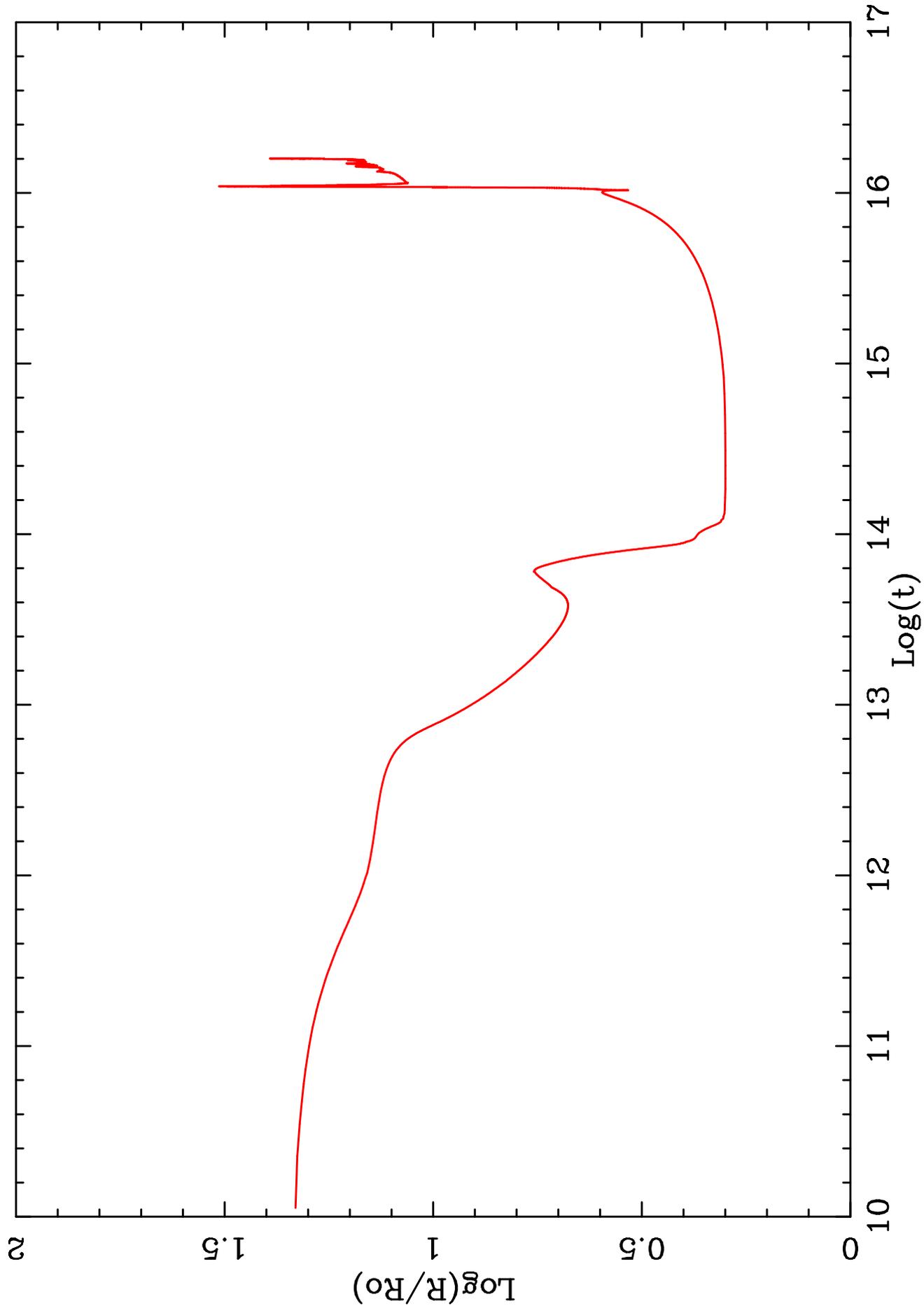
\end{document}